\input{aipcheck}

\documentclass[sort&compress,
    ,final            
  ]
  {aipproc}

\usepackage{version}

\layoutstyle{8x11double}

\begin{document}

\title{Confined  packings of frictionless spheres and polyhedra}


\author{Jean-Fran\c cois Camenen}{
  address={LUNAM, IFSTTAR, site de Nantes, Route de Bouaye CS4 44344 Bouguenais Cedex, France}
}

\author{Yannick Descantes}{
  address={LUNAM, IFSTTAR, site de Nantes, Route de Bouaye CS4 44344 Bouguenais Cedex, France}
}

\author{Patrick Richard}{
  address={LUNAM, IFSTTAR, site de Nantes, Route de Bouaye CS4 44344 Bouguenais Cedex, France}
  ,altaddress={Institut Physique de Rennes, Universit\'e de Rennes I, UMR CNRS 6251, 35042 Rennes, France} 
}

\begin{abstract}
By means of numerical simulations, we study the influence of confinement on  three-dimensional random close packed (RCP) granular materials subject to gravity. The effects of grain shape (spherical or polyhedral) and polydispersity on this dependence are investigated. 
{In agreement with a {simple} geometrical model,} the solid fraction is found to decrease linearly for increasing confinement no matter the grain shape. This decrease remains valid for bidisperse sphere packings although the gradient seems to reduce significantly when the proportion of small particles reaches $40\%$ by volume. 
The aforementioned model is extended to capture the effect of the confinement on the coordination number.
\end{abstract}

\maketitle


\section{Introduction}\label{sec:intro}
Real granular systems have boundaries that, for the sake of simplicity, are often neglected by scientists who  consider the system as infinite. 
This {assumption} is not always justified since boundaries modify the {system} local arrangement in their vicinity. Moreover, due to the steric hindrance of granular materials those structure modifications may propagate over distances of the order of several grain sizes. As a consequence, the behavior of granular systems may be strongly influenced by the presence of sidewalls even if the confinement length is large compared to the grain size~\cite{Richard_PRL_2008}.
Here, we focus on the geometric effect of the presence of sidewalls on the solid fraction and the coordination numbers
of quasi-static dense frictionless granular packings. We compare numerical results with a simple geometric model~\cite{Verman_Nature_1946,Brown_Nature_1946,Combe_PhD_2001,Desmond2009}
 based on the following configuration: a packing of particles is confined between two parallel and flat walls separated by a gap $W$. This model assumes that such a confined system is made of two boundary layers (of thickness $h$) and a bulk region and that the solid fraction of the boundary layers, $\phi_{BL}$, is lower than that of the bulk 
{region}  $\phi_{bulk}$.
The total solid fraction is then given by
\begin{equation} \label{eqn:weeks} \phi=\phi_{bulk}-{C}/{W},
\mbox{ with } C=2h\left(\phi_{bulk}-\phi_{BL}\right).\end{equation}
In the latter equation, the three parameters of the {geometrical} model ($\phi_{bulk}$, $\phi_{BL}$ and $h$) probably depend on grain shape and packing polydispersity. 
In this article we first describe our simulations, then we investigate how the solid fraction and the coordination number are  modified by confinement and how these modifications are influenced by packing polydispersity and grain shape. Finally, we present our conclusions.

\section{Simulation methodology}\label{sec:simmet}


Discrete numerical simulations were performed using the contact
dynamics (CD) method~\cite{Moreau94}.
All the details on the implementation of that method for our study as well as relevant references can be found in~\cite{Camenen_PRE_2012}.

\begin{figure}[htbp]
\includegraphics*[width=1\columnwidth]{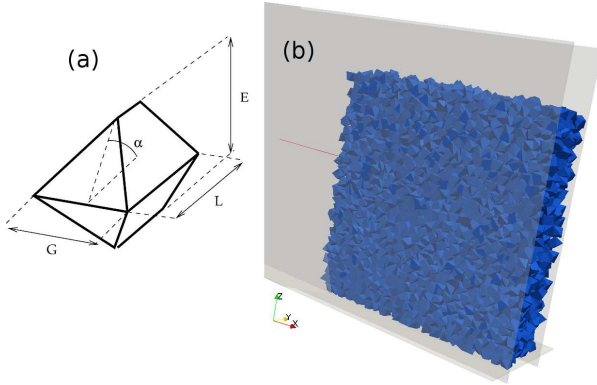} 
\caption{(a) Pinacoid, a model polyhedra characterized by its length $L$, width $G$, height $E$ and angle $\alpha$.
(b) Typical 3D snapshot of a pinacoid packing. 
}  \label{pinacoid}
\end{figure}

The simulated system is a three-dimensional dense assembly of $n$ frictionless rigid grains of mass density $\rho$, interacting with each other through totally inelastic collisions.
Two types of grains have been studied: spheres of average diameter $d$ and polyhedra of average characteristic dimension $d$. The polyhedra shape (Fig.~\ref{pinacoid}a) is that of a \emph{pinacoid}, with 8 vertices, 14 edges and 8 faces. It has three symmetry planes and is determined by four parameters: length $L$, width $G$, height $E$ and angle $\alpha$. According to an extensive experimental study with various rock types reported
by~\cite{Tourenq82}, the pinacoid gives the best fit among simple geometries for an aggregate grain. 
The pinacoid lengths
were taken identical ($L=G=E$) and equal to the characteristic dimension $d$
and angle $\alpha$ was set to $60^\circ$. 
For each grain shape, two grain diameters (or characteristic dimensions) have been considered: large $d_L$ and small $d_S = d_L/2$. 
The packing geometry is that of a parallelepiped (Fig.~\ref{pinacoid}b) of dimensions $L_x$ by $L_y$ by $L_z$. 
Periodic Boundary Conditions (PBC) 
{are} applied in the $x$ direction.
The packing is confined 
{in} the $y$ direction between two fixed parallel walls 
{separated by a $L_y=W$ large gap.}
In some cases, PBC are also applied along the $y$ axis to simulate unconfined reference state, with $W$ set to $20d_L$. 
The packing is supported on the $xy$-plane by a fixed frictionless bottom wall and delimited by a free surface at its top.
Grain samples are composed of various proportions of small and large grains having the same shape. 
Each population of grain size is randomly generated with a Gaussian distribution characterized by its mean $d$ and its variance $d^2/900$. Anyhow, for the sake of simplicity, packings made of a unique population of grains (either small or large) will be called "monodisperse", whereas packings made of small and large grains will be called "bidisperse". In the latter case, the proportions of small ($x_S$) and large ($x_L$) grains expressed as percentages by volume are of course linked through $x_S=100-x_L$.
A full description of the procedure used to build our packings is available in Ref.~\cite{Camenen_PRE_2012}.
The present study focuses on monodisperse sphere packings (MSP), bidisperse sphere packings (BSP) and monodisperse pinacoid packings (MPP). The gap between lateral walls $W$ takes discrete values between $5d_L$ and $20d_L$, and $L_x=20d_L$. 
The number $n$ of grains varies between 1,900 and 30,400 for spheres, depending on the proportion by volume of small grains, and between 3,600 and 15,000 for pinacoids.
All our packings are mechanically stable 
and homogeneous (see~\cite{Camenen_PRE_2012} for details). 
According to~\cite{roux_04}, random close-packed states of rigid frictionless grains (spherical or non-spherical) are equivalent to packing states in which the grains are homogeneously spread and in a stable equilibrium 
(\textit{i.e.} their potential energy is minimum) without crystallization or segregation. 
Furthermore, extensive investigation of the random close-packed state carried out by~\cite{Agnolin_PRE_2007} with spherical particles has evidenced the uniqueness of this state in the limit of infinitely large samples subject to fast isotropic~\cite{Camenen_PRE_2012} compression (to avoid cristallization). Hence, the influence of wall-induced confinement on the solid fraction and structure of dense packings may be assessed against the random close-packed state taken as the reference. Keeping in mind that our compaction method only allows to approximate the random close-packed state (our compression is not isotropic) and that the uniqueness of this reference state has only been evidenced {for} sphere packings, it is expected that meeting as much as possible the criteria stated by~\cite{roux_04} will lead to sufficiently repeatable solid fraction and microstructure characteristics for a given set of materials and system parameters to observe confinement effects for various grain shapes and polydispersity. 

\section{Solid fraction}\label{sec:solid_fraction}

\begin{figure}[htbp]
\includegraphics*[width=.9\columnwidth]{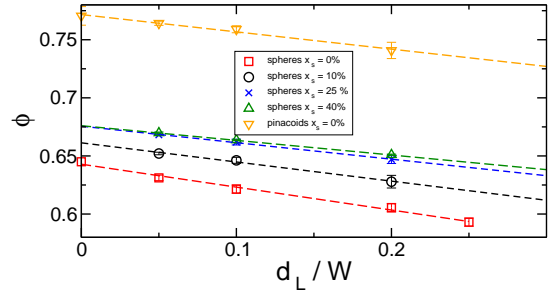}
\caption{Average solid fraction versus $d_L/W$ for MSP, BSP and MPP. The lines are fits from the 
model
[Eq.~\ref{eqn:weeks}]. 
}\label{fig:phi_vs_dsurW_MSP}
\end{figure}

Figure~\ref{fig:phi_vs_dsurW_MSP} reports the average solid fraction for BSP, MSP and MPP versus $d_L/W$. 
Each value is averaged over three simulations and error bars denote the corresponding standard deviation. A first expected observation is that, for a fixed $d_L/W$ value, an addition of small grains in a monosized sphere packing increases the solid fraction. 
More interestingly, an excellent agreement between our data and the {geometrical} model 
is found and the value of $\phi_{bulk}$ obtained for MSP is consistent with that of the random close packing ($0.64$). 
Note that in Ref.~\cite{Desmond2009}, the geometrical model has been compared with simulations of bidisperse sphere packing (50-50 binary mixture with particle size ratio of 1.4)  in the absence of gravity. Our results show that the validity of this model is much broader since it still holds in the presence of gravity for monodisperse sphere packings, for bidisperse sphere packings (independently of $x_s$) as well as for monodisperse pinacoid packings. This result is important in the framework of aggregates whose grains are far from being perfect spheres. 
Our results also show that when the fraction of small grains, $x_s$, increases, $C=2h (\phi_{bulk}-\phi_{BL})$) decreases. This can be the consequence of a decrease of the distance of propagation of the sidewall effects $h$ or/and of the difference $\phi_{Bulk}-\phi_{BL}$. 
To address this point we also study the local variation of the solid fraction close to the sidewalls. 
\begin{figure}[htbp]
\includegraphics*[width=.9\columnwidth]{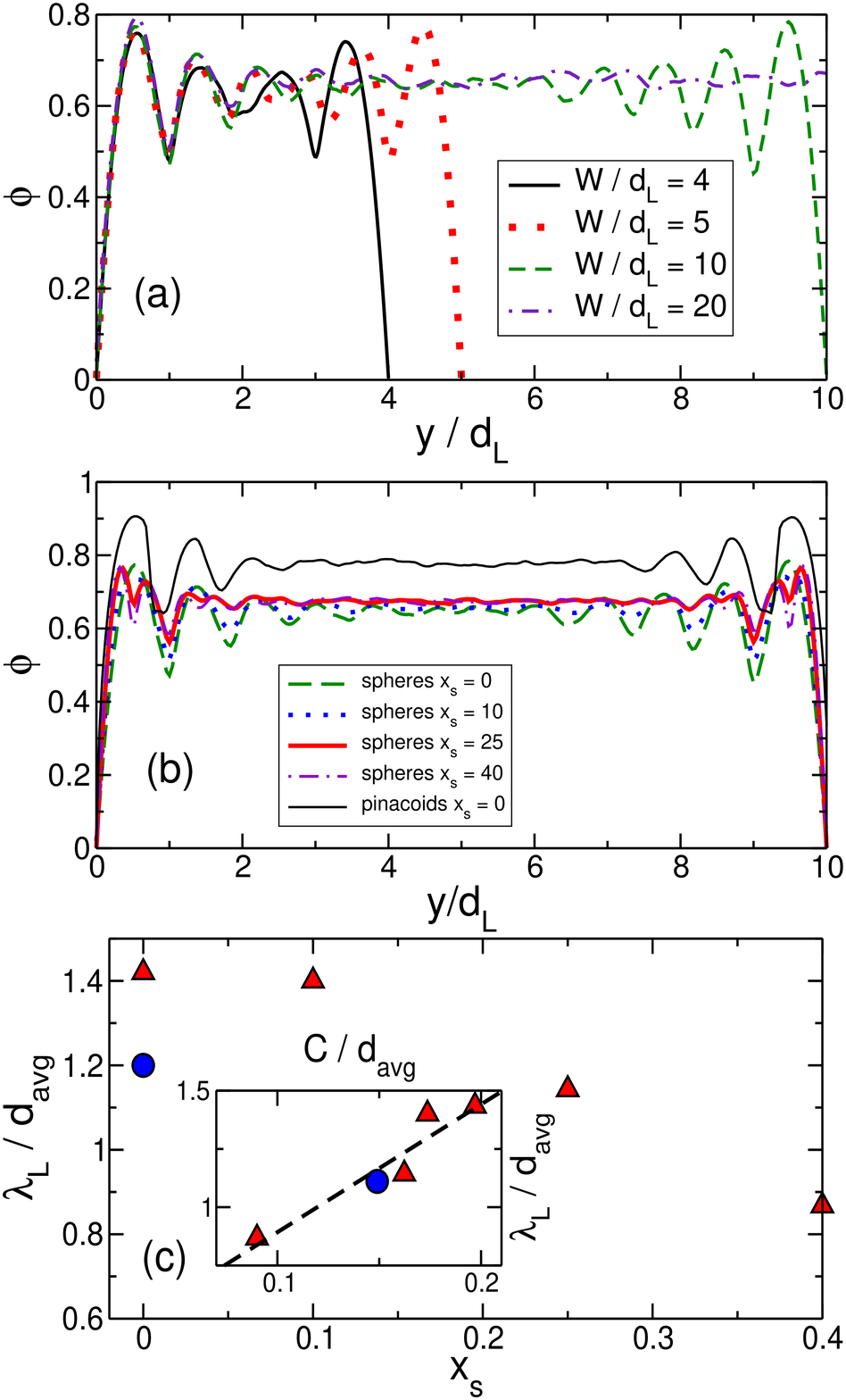}
\caption{Solid fraction profiles as a function of distance $y/d_L$ from confining wall for 
(a) MSP with $W=4d_L$, $5d_L$, $10d_L$ and $20d_L$, (b) BSP with $W=10d_L$ and $x_s = 0\%, 10\%, 25\%\mbox{ and }40\%$,
and  MPP with $W=10d_L$. (c) 
Characteristic length of confinement effect $\lambda_L$ versus $x_s$ for MSP and BSP (triangles) as well as for MPP (circles) with $W=10d$. 
The inset reports the same length versus that of the fit parameter in eq.~(\ref{eqn:weeks}): $C=2h(\phi_{bulk}-\Phi_{BL})$. The dashed line is a linear fit.
}\label{fig:Suzuki}
\end{figure}
Figure~\ref{fig:Suzuki} reports the solid fraction profile as a function of the distance $y/d_L$ to the left sidewall for MSP (a) and BSP and MPP (b). The solid fraction fluctuates with $y$
especially in the neighborhood of sidewalls and, if $W$ is large enough, it reaches a uniform value away from the sidewalls. 
{The aforementioned fluctuations 
{clearly reflect the} layering due to the presence of sidewalls
-- \textit{i.e.} an order propagation in the $y$ direction
~\cite{Suzuki2008}.}
For MSP, the confinement effect propagates over approximately $3d_L$ to $4d_L$. As a result, packings for which  $W<6d_L$ to $8d_L$ are influenced by the presence of walls over their full width and the order generated by the sidewalls propagates in the whole packing.
On the contrary for BSP as well as for MPP, the propagation seems to be shorter (approximately $1.5d_L$ to $2d_L$ for BSP and about $2d_L$ for MPP). 
{The presence of bidispersity or non-sphericity induces disorder 
{in} the vicinity of the sidewalls which 
{mitigates} the layering.}
We have shown in~\cite{Camenen_PRE_2012} that characteristic lengths of sidewall propagation for large ($\lambda_L$) and small ($\lambda_S$) grains can be extracted form those profiles.
{The values of $\lambda_L$ (those of $\lambda_S$ are not statistically relevant for $x_S < 0.25$), normalized by the average grain size $d_{avg}$}, obtained this way are reported in Fig.~\ref{fig:Suzuki}c for $W=10d_L$.
For sphere packings, that quantity is found to decrease when the fraction of small spheres $x_s$ increases. Indeed, for $x_s=0$ we have $\lambda_L/d_{avg}\approx 1.4$ whereas $\lambda_L/d_{avg}\approx 0.85$ for $x_s=40\%$. This decrease proves that the polidispersity mitigates {the} confinement effect. 
{Morover, the fact that 
$\lambda_L/d_{avg}$ decreases with $x_s$ demonstrates that 
{$\lambda_L$ decreases quicker than the mean grain size.}}
For MPP we obtain $\lambda/ d_{avg} = 1.2$ which is smaller than the value obtained for MSP. This indicates that the sidewall effect is also 
{mitigated by an increase in grain angularity}.
Hence, 
characteristic length $\lambda_{{L}}$ is expected to correlate with the 
 {thickness} $h$  {of the boundary layers} (cf. Eq.~\ref{eqn:weeks}). 
 In the inset of Fig.~\ref{fig:Suzuki}c we report $\lambda_{{L}}$ versus $C{=2h(\phi_{bulk}-\phi_{BL})}$ and observe a good linear correlation between these two parameters. Furthermore, the data for both sphere and pinacoid packings 
 collapse on the same straight line whose intercept is equal to zero.\\

\section{Coordination number} \label{sec:Z}

\begin{figure}[htbp]
\includegraphics*[width=.9\columnwidth]{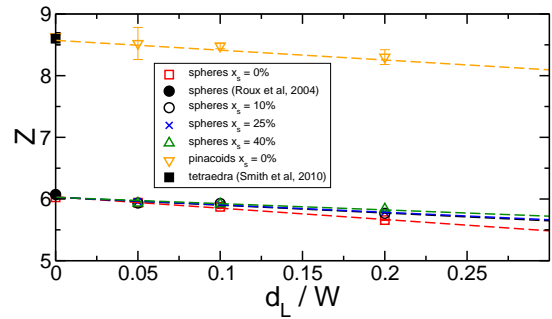}
\caption{Coordination number versus $d_L/W$ for MSP, BSP and MPP. 
}\label{fig:weeks_coord}
\end{figure}

Figure~\ref{fig:weeks_coord} shows the variations of the coordination number with $d_L/W$ for MSP, BSP and MPP. 
Each value is averaged over three simulations and the error bars denote the corresponding standard deviation. Preliminary examination of our results obtained with bi-periodic boundary conditions 
suggests the following remarks: for sphere packings, the calculated coordination number is $6.027\pm{0.012}$, which is very close to the $6.073\pm{0.004}$ value calculated by~\cite{roux_04} in the RCP state. For pinacoid packings, the calculated coordination number is $8.581\pm{0.068}$. 
Such a high coordination number value has already been observed in disordered packings of particles having a similar shape~\cite{Smith_PRE_2010}. 
When confinement increases, the coordination number decreases linearly for both MSP and MPP, which is consistent with the linear decrease of the solid fraction evidenced in Fig.~\ref{fig:phi_vs_dsurW_MSP}. 
The aforementioned linear relation between $Z$ and $1/W$ suggests a generalization 
of {the}  {geometrical} model to the coordination number. For this purpose, let us define $Z_{bulk}$ and $Z_{BL}$, respectively the coordination number for the bulk region and the coordination number for the boundary layers. Writing the coordination number as the average of $Z_{bulk}$ and $Z_{BL}$ weighted by the thicknesses of their respective zones ($W-2h_Z$ and $2h_Z$) lead to
$Z=Z_{bulk}-{C_Z}/{W},
\mbox{ with }C_Z=2h_Z(Z_{bulk}-Z_{BL}).$ 
%
Figure~\ref{fig:weeks_coord} also shows that the influence of polydispersity on packing coordination number $Z$ decreases to zero when the confinement diminishes, which is an expected result. Indeed, in the unconfined state, the lack of contacts of small spheres  {with others (due to the steric hindrance of large ones)} 
 is compensated by the excess of contacts of large spheres  {with small ones~\cite{Roux_BLCPC_2007}.}

\begin{figure}[htbp]
\includegraphics*[width=.9\columnwidth]{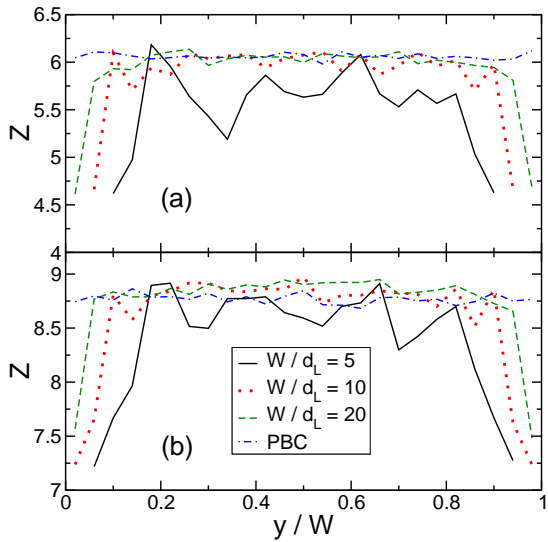}
\caption{Coordination number profiles (along $y$) for MSP (a) and for MPP (b) for several gap widths. These profiles evidence a constant central zone and two drop zones close to the sidewalls.}\label{fig:prof-Z}
\end{figure}

To investigate the coordination number decrease with increasing confinement, Fig.~\ref{fig:prof-Z} reports coordination number profiles (averaged over 3 simulations) in the $y$ direction (normal to the sidewalls) for sphere and for pinacoid packings. 
In confined state, all these profiles evidence a central zone where the coordination number is almost unchanged compared to the unconfined reference state (except for sphere packings with $W=5d_L$), and two "drop zones" in contact with the sidewalls where the coordination number symmetrically drops by $1.4$ (spheres) to $1.5$ contacts (pinacoids) from their respective unconfined reference state. The thicknesses of these drop zones look identical to that of the boundary layers described 
in  {the}   {geometrical model}~\cite{Desmond2009}, leading to the same conclusion that grain angularity mitigates the effect of 
 sidewalls  {on the coordination number drop in their vicinity}.
 This observation is confirmed by comparing $\zeta_\phi=C/\phi_{bulk}$ with $\zeta_Z=C_Z/Z_{bulk}$. Those two quantities are comparable for MSP, BSP and MPP showing that
the propagation of the confinement effect  is comparable for 
$\phi$ and $Z$. 

Similarly to what is observed for the solid fraction an increasing polydispersity does not seem to impact coordination number in the bulk {region}, but reduces the thickness of the boundary layers, hence mitigates the effect of sidewalls confinement on the coordination number~\cite{Camenen_PRE_2012}.
%
%
%
%
%
%



\section{Conclusion}\label{sec:conclusion}
In this work, we have shown how a confining boundary alters the solid fraction as well as the coodination number of static {frictionless}
granular materials {compacted under their own weight using}  
 the non-smooth 
 contact dynamics {simulation} method. We did not restrict ourselves to sphere packings but extended our work to packings made of a particular type of polyhedra: pinacoids.\\
We have shown that the confinement effect is lowered by both the polydispersity and the angularity of grains. 
We also  have demonstrated that the  {geometrical model ~\cite{Verman_Nature_1946,Brown_Nature_1946,Combe_PhD_2001,Desmond2009}} that captures the linear evolution of the solid fraction versus $-1/W$  is valid for sphere packings and for pinacoid packings and that it holds whatever the packing polydispersity.
Interestingly, this model, initially derived for the packing fraction can be extended to capture the effect of confinement on the coordination number. The characteristic length quantifying the effect of the sidewalls is found to be the same for those two quantities.
Several perspectives arise from this study, among which the need to investigate the contact types in relation with the force network.


\end{document}